\documentclass[12pt,reqno,fleqn]{amsart}
\usepackage{latexsym}
\usepackage{amssymb}
\usepackage{amsxtra}
\usepackage{amsmath}
\usepackage{amsfonts}
\usepackage[dvips]{graphicx}
\usepackage{amsmath}
\usepackage{amssymb,amsmath,amsthm,amscd,mathrsfs,amsbsy,amsfonts}
\usepackage{pstricks,pst-node}
\usepackage{amsbsy,young,colortbl,cite}

\usepackage{graphicx}
\usepackage[metapost]{mfpic}
\usepackage{mflogo}

\usepackage{epstopdf}
\usepackage{graphics}

\def\@makecaption#1#2{\vskip\abovecaptionskip
  \sbox\@tempboxa{\small #1: #2}%
  \ifdim \wd\@tempboxa >\hsize \small #1: #2\par
  \else \global \@minipagefalse \hb@xt@\hsize{\hfil\box\@tempboxa\hfil}\fi
  \vskip\belowcaptionskip}
\newcommand{\cleqn}{\setcounter{equation}{0}}
\newcommand{\clth}{\setcounter{theorem}{0}}
\newcommand {\sectionnew}[1]{\section{#1}\cleqn\clth}
\newtheorem{theorem}{Theorem}[section]

\newcommand{\T}{\mathbb{T}}

\makeatletter
    
    \newcommand{\Rmnum}[1]{\expandafter\@slowromancap\romannumeral #1@}
  \makeatother

\def\({\left(}
\def\){\right)}
\def\[{\begin{eqnarray}}
\def\]{\end{eqnarray}}
\def\d{\partial}

\def\d{\partial}
\def\ep{\epsilon}
\def\La{\Lambda}
\def\la{\lambda}

\def\om{\omega}
\def\De{\Delta}
\begin{document}

\title{Darboux transformation and positons of the inhomogeneous Hirota and the Maxwell-Bloch equation }
\author{Chuanzhong Li\dag
, Jingsong He
\ddag}

\dedicatory {\small Department of Mathematics,  Ningbo University, Ningbo, 315211, China\\
\dag lichuanzhong@nbu.edu.cn\\
\ddag hejingsong@nbu.edu.cn}
\thanks{\ddag Corresponding author}
\texttt{}

\date{}

\begin{abstract}
 In this paper, we derive Darboux transformation of the inhomogeneous Hirota and the Maxwell-Bloch(IH-MB) equations which is governed by femtosecond pulse propagation through inhomogeneous doped fibre. The determinant representation of Darboux transformation is used to derive  soliton solutions, positon solutions of the IH-MB equations.

\end{abstract}

\maketitle
\tableofcontents
\allowdisplaybreaks
 \setcounter{section}{0}

\sectionnew{Introduction }

In recent years, nonlinear science has emerged as a powerful subject for explaining the mystery present in the challenges of science and technology today.
Among nonlinear science, the interplay between dispersion and nonlinearity gives rise
to several important phenomena in optical fibers, including
parametric amplification, wavelength conversion, modulational
instability(MI), soliton propagation and so on.
Among all concepts,  solitons, positons and rogons have  been not only the subject of intensive
research in oceanography\cite{OCR,Osborne} but
also it has been studied extensively in several areas, such as  Bose-Einstein condensate,  plasma, superfluid, finance, optics and so on \cite{liuwuming,liuwuming2,liuwuming3,liuwuming4,shuweiJPA,shuweiJMP,nlshighrogue}.

An important ingredient in the development of the theory of soliton and of complete integrability has been the interplay between mathematics and physics. In 1973, Hasegawa and Tappert \cite{AB1} modeled the propagation of coherent optical pulses in optical fibres by nonlinear Schr\"odinger (NLS) equation without the inclusion of fibre loss. They showed theoretically that generation and propagation of shape-preserving pulses called solitons in optical fibres is possible by balancing the dispersion and nonlinearity.

In 1967, McCall and Hahn \cite{McCallPRL} had explained a special type of lossless pulse propagation in two-level resonant media. They have discovered the self-induced transparency(SIT) which can be explained by Maxwell-Bloch(MB) equations. The coherent absorption takes place and the media becomes optically transparent to that particular
wavelength when the energy difference between the two levels of the media coincide with the optical wavelength. Burtsev and Gabitov\cite{Burtsev} have considered MB equations with pumping and damping
which is useful in optical pumping during the propagation of optical pulses in resonant atoms, and in their paper the Lax pair was presented for deformed MB systems.

The important constraint to the NLS soliton namely the optical losses can be somewhat compensated with the effect of SIT. Therefore the system will be governed by the coupled system of the NLS equation and the MB equation (NLS-MB equations) if we consider these effects for a large width pulse. The coexistence of NLS solitons and SIT solitons in erbium-doped resonant fibres was experimentally observed by Nakazawa et al \cite{Nakazawa,Nakazawa2}. Recently, multi-soliton solutions of coupled NLS-MB equation was shown in \cite{NLSMBPorsezian}. In \cite{heTheor}, the periodic solutions have been generated through Darboux transformation and later rogue wave solutions were derived from breather solution in \cite{hexuNLSMB}.  Modeling photonic crystal fiber for efficient soliton pulse propagation at 850 nm was surveyed in \cite{por850}. It presents new types of Dark-in-the-Bright solutions also called dipole soliton for the higher order nonlinear
NLS (HNLS) equation with non-Kerr nonlinearity under some parametric conditions and subject to constraint
relation among the parameters in optical context in \cite{Porsezian2012}. Impact of fourth-order dispersion in the
modulational instability spectra of wave
propagation in glass fibers with saturable
nonlinearity was considered in \cite{PorsezianMI}.
The HNLS-MB equations as a higher-order correction of NLS-MB equations were  shown that they allow soliton-type pulse propagation under a particular parametric condition\cite{Nakkeeran95,Nakkeeran95jmodopt}.

For a reduced dynamical equation, the erbium-doped fibre system was proven to allow soliton-type pulse propagation with pumping\cite{Nakkeeranjpa}. The Lax pair and the exact soliton
solution for Higher-order nonlinear Schr\"odinger and Maxwell-Bloch(HNLS-MB) equations with pumping was derived in \cite{Nakkeeranjpa}.

Kodama\cite{Kodama} has shown that with suitable transformation and omitting the higher-order terms, higher order Nonlinear Schr\"odinger equation equation can be reduced to the Hirota equation\cite{hirotaeq}whose rogue wave solution is already reported  in \cite{Akhmedievhirota,taohirota}. In a similar way,after suitable choice of self steepening and self frequency effects, the                                                                                                                                                                                                                                                                                HNLS-MB equations can be reduced to a coupled system of the Hirota equation and MB equation\cite{PorsezianPRL}.
The H-MB equations can be seen as the higher order correction of the NLS-MB equations and is the coupled system of the Hirota equation
and the MB equation\cite{rogueHMB}. The H-MB system has been shown to be integrable and also admits the Lax pair and other required properties for complete integrability \cite{PorsezianPRL}.

It is well known that the Darboux transformation is an efficient method to generate the soliton solutions for integrable equations\cite{Matveev,guoboling}. The determinant representation of n-fold Darboux transformation of AKNS system was given in \cite{Hedeterminant}. In \cite{rogueHMB}, it constructed n-folds Darboux transformation of the H-MB equations, meanwhile the rogue wave solutions of the H-MB equations were obtained using the Darboux transformation.

Inhomogeneous integrable equations become more and more attractive\cite{yanzhenya}.
Recently, K. Porsezian and C. G. Latchio Tiofack, Thierry B. Ekogo, etal. consider dynamics of bright solitons and their collisions for
the inhomogeneous coupled nonlinear Schr\"odinger-Maxwell-Bloch equations which describes propagation of an
optical soliton in an inhomogeneous nonlinear waveguide doped with two level
resonant atoms\cite{PorsezianNLSinhomogeneous}.
Soliton interactions in a generalized inhomogeneous Hirota-Maxwell-Bloch(IH-MB) system were considered in \cite{zhangjiefang,BoTian} with symbolic computation but positon solutions of IH-MB is still unknown.

 The purpose of this paper is to derive the determinant representation of Darboux transformation which is used to derive  soliton solutions, positon solutions of the IH-MB equations.

The paper is organized as follows.  In Section 2, the Lax representation of IH-MB equations will be introduced firstly.  In Section 3, we derived the one-fold Darboux transformation of the H-MB equations. In Section 4, the determinant-formed generalization of one-fold Darboux transformation to 2-fold Darboux transformation of the IH-MB equations will be given.  Using these Darboux transformations, one soliton, two soliton and positons are derived in Section 5 and Section 6 by assuming trivial seed solutions.  Section 7 is devoted to conclusion and discussions.

\sectionnew{ Lax representation of the IH-MB system   }
 In this paper, we will concentrate on the inhomogeneous Hirota and the Maxwell-Bloch(H-MB) system  as following specific form\cite{BoTian,zhangjiefang},

\[
E_z&=&-(a_1(z)E_t+a_2(z)E+ia_3(z)E_{tt}+a_4(z)E_{ttt}+a_5(z)|E|^2E_t\\
\notag &&+ia_6(z)|E|^2E+a_7(z)p),\\
p_t&=&2b_1(z)E \eta-2ib_2(z)\om p,\\
\eta_t&=&-b_1(z)(Ep^*+E^* p),
\]
with constraint
\[a_2=\frac{\frac{\d b_1}{\d z}}{b_1};a_5=6a_4b_1^2;a_6=2a_3b_1^2.\]
In the equations above, $z$ and $t$  represent the normalized
distance and time respectively, $E(z, t)$ denotes the slowly varying envelope
axial field, $p(z, t)$ is the measure of the
polarization of the resonant medium, and $\eta(z, t)$ represents
the extent of the population inversion. $a_1(z)$
results from the group velocity and $a_2(z)$ describes
the amplification or absorption. The coefficients
$a_3(z) \sim a_6(z)$ represent the group velocity
dispersion (GVD), the
third-order dispersion(TOD)\cite{TOD}, self-steepening
(SS)\cite{SS}, and self-phase modulation respectively.
$a_7(z)$ is the parameter describing the averaging
with respect to inhomogeneous broadening of the resonant
frequency. $b_1(z)$ and $b_2(z)$ depict the character
of interactions between the propagation field and the
erbium atoms. The real parameter $\omega$ is a constant corresponding
to the frequency, and the $*$ denotes
the complex conjugate.
If we set
\[a_1=a_2=0, a_3=-\frac12\alpha,a_4=-\beta,a_5=-6\beta,a_6=-\alpha,a_7=-2,b_1=1,b_2=-1,\] the inhomogeneous H-MB equation will be reduced to
H-MB equation as following

\[
E_z&=&i\alpha(\frac12 E_{tt}+|E|^2E)+\beta(E_{ttt}+6|E|^2E_t)+2p,\\
p_t&=&2i\om p+2E \eta,\\
\eta_t&=&-(Ep^*+E^* p).
\]

We will call the inhomogeneous Hirota and the Maxwell-Bloch system when $\alpha=2, \beta=-1$ the classical H-MB equation.
The linear eigenvalue problem of IH-MB takes the form

\[\label{linear}
\Phi_t&=&U\Phi,\\ \label{linear2}
\Phi_z&=&V\Phi,
\]
where $U$ and $V$ can be expressed in following polynomials about complex constant eigenvalue parameter $\lambda$
\[
U&=&\la\left(\begin{matrix}1& 0\\ 0& -1
\end{matrix}\right)+\left(\begin{matrix}0& b_1(z)E\\ -b_1(z)E^*& 0
\end{matrix}\right)=\la \sigma_3+U_0,\\ \notag
V&=&\la^3\left(\begin{matrix}-4a_4& 0\\ 0& 4a_4
\end{matrix}\right)+\la^2\left(\begin{matrix}-2ia_3& -4a_4b_1E\\ 4a_4b_1E^*& 2 ia_3
\end{matrix}\right)+\la\left(\begin{matrix}-a_1-2a_4b_1^2 |E|^2 &-2ib_1(a_3E-ia_4E_t)\\ 2ia_3b_1E^*-2a_4b_1E_t^*&a_1+2a_4b_1^2 |E|^2
\end{matrix}\right)\\ \notag
&&+\left(\begin{matrix}-C_1 &-b_1B_1\\ -b_1B_1^*& C_1
\end{matrix}\right)+A_1\left(\begin{matrix}\eta &-p\\ -p^*& -\eta
\end{matrix}\right)\\
&=&\la^3V_3+\la^2V_2+\la V_1+V_0+\frac{1}{\la+i\om b_2}V_{-1},
\]
where
\[V_{-1}=-\frac{b_1a_7}{2}\left(\begin{matrix}\eta &-p\\ -p^*& -\eta
\end{matrix}\right),\]
\[
A_1£º=-\frac{b_1a_7}{2(\la+i\om b_2)},\ \ B_1:=2a_4b_1^2E^*E^2+a_1E+ia_3E_t+a_4E_{tt},\]
\[C_1:=ia_3b_1^2EE^*+a_4b_1^2(E^*E_t-EE_t^*).\]
$V_i$ denotes the coefficient matrix of term $\la^i$
and
\[\Phi &= &\Phi(\lambda)=\left(\begin{matrix}
\Phi_1(\lambda,t,z)\\
\Phi_2(\lambda,t,z)
\end{matrix}\right)
\] is an eigenfunction associated with eigenvalue parameter $\lambda$ of  linear system
eq.(\ref{linear}-\ref{linear2}).

Using the linear equations of H-MB equations, One-fold Daroux transformation for IH-MB equation will be introduced in the next section.

\sectionnew{ One-fold Daroux transformation for the IH-MB equation  }

In this section, we will give the detailed proof of the one-fold Daroux transformation for the IH-MB equation.
Firstly, we consider the transformation about linear function $\Phi$
\[
\Phi'&=&T\Phi=(\la A-S)\Phi,
\]
where
\[A=\left(\begin{matrix}a_{11}& a_{12}\\ a_{21}& a_{22}
\end{matrix}\right),\ \ \
S=\left(\begin{matrix}s_{11}& s_{12}\\ s_{21}& s_{22}
\end{matrix}\right).
\]

New function $\Phi'$ is supposed to satisfy
\[
\Phi'_t&=&U'\Phi',\\
\Phi'_z&=&V'\Phi'.
\]

Then matrix $T$ can be proven to satisfy following identities
\[\label{tequation}
T_t+TU&=&U'T,\\ \label{zequation}
T_z+TV&=&V'T.
\]
Bring the form of matrices $A$ and $S$ into eq.\eqref{tequation} and comparing the coefficients of both sides will lead to following condition
\[a_{12}=a_{21}=0,\ \ \ (a_{11})_t=(a_{22})_t=0.\]
Therefore we will  choose $A=I$ and $T=\la I-S$ in the following part of this paper.
 The relation between $E,p,\eta$ and new solutions $E',p',\eta'$ which is called Darboux transformation can be got by eq. \eqref{tequation} and eq. \eqref{zequation}.

From \eqref{tequation}, we have
\[\label{E'andE}
E'&=&E+2b_1^{-1}s_{12},\\
S_t&=&\left(\begin{matrix}0& b_1E\\ -b_1E^*& 0
\end{matrix}\right)S-S\left(\begin{matrix}0& b_1E\\ -b_1E^*& 0
\end{matrix}\right)-[S,\sigma_3]S,
\]
and $S$ should have a condition as $s_{21}=s^*_{12}.$
By \eqref{zequation}, following identity can be got
\[\label{zeqdetailed}
&&-S_z+(\la I-S)(\la^3V_3+\la^2V_2+\la V_1+V_0+\frac{1}{\la+i\om b_2}V_{-1})\\
&=&(\la^3V_3+\la^2V'_2+\la V'_1+V'_0+\frac{1}{\la+i\om b_2}V'_{-1})(\la I-S).
\]
Multiplying both sides of eq.\eqref{zeqdetailed} by $\la I-S$ can lead to
\begin{eqnarray*}
&&-S_z(\la+i\om b_2)+(\la I-S)(\la^3(\la+i\om b_2)V_3+\la^2(\la+i\om b_2)V_2+\la(\la+i\om b_2) V_1+V_0(\la+i\om b_2)+V_{-1})\\
&=&(\la^3(\la+i\om b_2)V_3+\la^2(\la+i\om b_2)V'_2+\la (\la+i\om b_2) V'_1+V'_0(\la+i\om b_2)+V'_{-1})(\la I-S).
\end{eqnarray*}

 For term with $\la^0$, we get following identity
 \begin{eqnarray*}
S_zi\om b_2+S(i\om b_2 V_0+V_{-1})
=(i\om b_2 V'_0+V'_{-1})S,
\end{eqnarray*}
which further leads to
\begin{eqnarray*}
S_z
&=&( V'_0-i\om^{-1} b_2^{-1}V'_{-1})S-S(-i\om^{-1} b_2^{-1}V_{-1}+V_0).
\end{eqnarray*}
For term with $\la$, we get following identity

\begin{eqnarray*}
S_z
&=&(i\om b_2 V_0+V_{-1})
-S(i\om b_2 V_1+V_0)+(i\om b_2 V'_1+V'_0)S+(-i\om b_2 V'_0-V'_{-1}).
\end{eqnarray*}
For term with $\la^2$, we get following identity
\begin{eqnarray*}
(i\om b_2 V_1+V_0)-S(i\om b_2 V_2+V_1)
&=&(i\om b_2 V'_1+V'_0)-(i\om b_2 V'_2+V'_1)S.
\end{eqnarray*}
For term with $\la^3$, we get following identity
\begin{eqnarray*}
(i\om b_2 V_2+V_1)-S(i\om b_2 V_3+V_2)
&=&(i\om b_2 V'_2+V'_1)-(i\om b_2 V_3+V'_2)S.
\end{eqnarray*}
For term with $\la^4$, we get following identity

\begin{eqnarray*}
V'_2=V_2-[S,V_3],
\end{eqnarray*}
From above several identities, we can get
\[\label{E'andE2}
E'&=&E+2b_1^{-1}s_{12},
\]
\begin{eqnarray}\label{V'-1dressing}
V'_{-1}=(S+i\om b_2 ) V_{-1}(S+i\om b_2)^{-1},
\end{eqnarray}
which gives one fold transformation of  one-fold Darboux transformation of H-MB equations.

Suppose
\[\label{SandH}
S=H\La H^{-1}\]
where
$\La=\left(\begin{matrix}\la_1& 0\\ 0& \la_2
\end{matrix}\right)$, $H=\left(\begin{matrix}\Phi_1(\la_1,t,z)& \Phi_1(\la_2,t,z)\\ \Phi_2(\la_1,t,z)& \Phi_2(\la_2,t,z)
\end{matrix}\right):=\left(\begin{matrix}\Phi_{1,1}& \Phi_{1,2}\\ \Phi_{2,1}& \Phi_{2,2}
\end{matrix}\right).$

In order to satisfy the constraints of $S$ and make $V'_{-1}$ having similar form as $V_{-1}$, i.e. $s_{21}=s^*_{12},$
following constraint will be considered
\[
\la_2&=&-\la_1^*,\ \ s_{11}=-s^*_{22},\\\label{H}
H&=&\left(\begin{matrix}\Phi_1(\la_1,t,z)& -\Phi^*_2(\la_1,t,z)\\ \Phi_2(\la_1,t,z)& \Phi^*_1(\la_1,t,z)
\end{matrix}\right).\]

The detailed determinant form of one-fold Darboux transformation of IH-MB equations in form of eigenfunctions will be given in the next section.

\sectionnew{Determinant representation of Darboux transformation}

In this section, we will give determinant representation of the first two Darboux transformation of the IH-MB equations. Other higher-order Darboux transformation can be got in similar way which can be seen clearly in our paper \cite{rogueHMB}.
Firstly, we introduce n eigenfunctions $\left(\begin{matrix}\Phi_{1,i}\\\Phi_{2,i}\end{matrix}\right)=\Phi(\la=\la_i), i=1,2$
with  constraints on eigenvalues as $\la_{2i-1}=-\la_{2i}^*$ and  the reduction conditions on eigenfunctions as $\Phi_{2,2i}=\Phi_{1,2i-1}^*,\ \ \Phi_{2,2i-1}=-\Phi_{1,2i}^*$.

As the simplest Darboux transformation, the determinant representation of one-fold Darboux transformation  of the IH-MB equations will be given in the following theorem.
\begin{theorem}
The one-fold  Darboux transformation of the IH-MB equations is as following
\[T_1(\la,\la_1,\la_2)=\la I+t_0^{[1]}=\frac{1}{\De_1}\left(\begin{matrix}(\T_1)_{11}&(\T_1)_{12}\\
(\T_1)_{21}&(\T_1)_{22}
\end{matrix}\right)\]
where

\[t_0^{[1]}=\frac{1}{\De_1}\left(\begin{matrix}\left|\begin{matrix}
\Phi_{2,1}&\la_1\Phi_{1,1}\\
\Phi_{2,2}&\la_2\Phi_{1,2}
\end{matrix}\right|&-\left|\begin{matrix}
\Phi_{1,1}&\la_1\Phi_{1,1}\\
\Phi_{1,2}&\la_2\Phi_{1,2}
\end{matrix}\right|\\
\\
\left|\begin{matrix}
\Phi_{2,1}&\la_1\Phi_{2,1}\\
\Phi_{2,2}&\la_2\Phi_{2,2}
\end{matrix}\right|&-\left|\begin{matrix}
\Phi_{1,1}&\la_1\Phi_{2,1}\\
\Phi_{1,2}&\la_2\Phi_{2,2}
\end{matrix}\right|
\end{matrix}\right),\ \ \De_1=\left|\begin{matrix}
\Phi_{1,1}&\Phi_{2,1}\\
\Phi_{1,2}&\Phi_{2,2}
\end{matrix}\right|, \]

\[(\T_1)_{11}&=&\left|\begin{matrix}1&0&\la\\
\Phi_{1,1}&\Phi_{2,1}&\la_1\Phi_{1,1}\\
\Phi_{1,2}&\Phi_{2,2}&\la_2\Phi_{1,2}\
\end{matrix}\right|,\ \ (\T_1)_{12}=\left|\begin{matrix}0&1&0\\
\Phi_{1,1}&\Phi_{2,1}&\la_1\Phi_{1,1}\\
\Phi_{1,2}&\Phi_{2,2}&\la_2\Phi_{1,2}
\end{matrix}\right|,\\
(\T_1)_{21}&=&\left|\begin{matrix}1&0&0\\
\Phi_{1,1}&\Phi_{2,1}&\la_1\Phi_{2,1}\\
\Phi_{1,2}&\Phi_{2,2}&\la_2\Phi_{2,2}
\end{matrix}\right|,\ \ (\T_1)_{22}=\left|\begin{matrix}0&1&\la\\
\Phi_{1,1}&\Phi_{2,1}&\la_1\Phi_{2,1}\\
\Phi_{1,2}&\Phi_{2,2}&\la_2\Phi_{2,2}
\end{matrix}\right|,\]
and
\[U^{[1]}=U+[T_1,\sigma_3],\ \ V^{[1]}_{-1}=T_1|_{\la=-i\om b_2}V_{-1}T_1^{-1}|_{\la=-i\om b_2},\]
\[\label{1darbouxE}E^{[1]}&=&E+2b_1^{-1}s_{12}=E-2b_1^{-1}\frac{(T_1)_{12}}{\De_1},\\ \label{1darbouxp}
p^{[1]}&=&\frac{2\eta(T_1)_{11}(T_1)_{12}-p^*(T_1)_{12}(T_1)_{12}+p(T_1)_{11}(T_1)_{11}}{(T_1)_{11}(T_1)_{22}-(T_1)_{12}(T_1)_{21}}|_{\la=-i\om b_2},\\ \label{1darbouxy}
\eta^{[1]}&=&\frac{\eta((T_1)_{11}(T_1)_{22}+(T_1)_{12}(T_1)_{21})-p^*(T_1)_{12}(T_1)_{22}+p(T_1)_{11}(T_1)_{21}}{(T_1)_{11}(T_1)_{22}-(T_1)_{12}(T_1)_{21}}|_{\la=-i\om b_2}.
\]

\end{theorem}
We can find the transformation $T_1$ has following property
 \[T_1(\la;\la_1,\la_2)|_{\la=\la_i}\left(\begin{matrix}\Phi_{1,i}\\
\Phi_{2,i}
\end{matrix}\right)=0,\]
where $i=1,2.$

This one-fold transformation will be used to generate one-soliton solution from trivial seed solution of the IH-MB equation.

In the next part, we will generalize the Darboux transformation to two-fold case which is contained in the following theorem.
\begin{theorem}
The two-fold  Darboux transformation of H-MB equation is as following
\[T_2(\la;\la_1,\la_2,\la_3,\la_4)=\la^2 I+t_1^{[2]}\la+t_0^{[2]}=\frac{1}{\De_2}\left(\begin{matrix}(\T_2)_{11}&(\T_2)_{12}\\
(\T_2)_{21}&(\T_2)_{22}
\end{matrix}\right)\]
where

\[\De_2&=&
\left|\begin{matrix}\Phi_{1,1}&\Phi_{2,1}&\la_1\Phi_{1,1}&\la_1\Phi_{2,1}\\
\Phi_{1,2}&\Phi_{2,2}&\la_2\Phi_{1,2}&\la_2\Phi_{2,2}\\
\Phi_{1,3}&\Phi_{2,3}&\la_3\Phi_{1,3}&\la_3\Phi_{2,3}\\
\Phi_{1,4}&\Phi_{2,4}&\la_4\Phi_{1,4}&\la_4\Phi_{2,4}
\end{matrix}\right|,\]

\[\notag (\T_2)_{11}&=&\left|\begin{matrix}1&0&\la&0&\la^2\\
\Phi_{1,1}&\Phi_{2,1}&\la_1\Phi_{1,1}&\la_1\Phi_{2,1}&\la_1^2\Phi_{1,1}\\
\Phi_{1,2}&\Phi_{2,2}&\la_2\Phi_{1,2}&\la_2\Phi_{2,2}&\la_2^2\Phi_{1,2}\\
\Phi_{1,3}&\Phi_{2,3}&\la_3\Phi_{1,3}&\la_3\Phi_{2,3}&\la_3^2\Phi_{1,3}\\
\Phi_{1,4}&\Phi_{2,4}&\la_4\Phi_{1,4}&\la_4\Phi_{2,4}&\la_4^2\Phi_{1,4}
\end{matrix}\right|,\ \ (\T_2)_{12}=\left|\begin{matrix}0&1&0&\la&0\\
\Phi_{1,1}&\Phi_{2,1}&\la_1\Phi_{1,1}&\la_1\Phi_{2,1}&\la_1^2\Phi_{1,1}\\
\Phi_{1,2}&\Phi_{2,2}&\la_2\Phi_{1,2}&\la_2\Phi_{2,2}&\la_2^2\Phi_{1,2}\\
\Phi_{1,3}&\Phi_{2,3}&\la_3\Phi_{1,3}&\la_3\Phi_{2,3}&\la_3^2\Phi_{1,3}\\
\Phi_{1,4}&\Phi_{2,4}&\la_4\Phi_{1,4}&\la_4\Phi_{2,4}&\la_4^2\Phi_{1,4}
\end{matrix}\right|,\\
\notag(\T_2)_{21}&=&\left|\begin{matrix}1&0&\la&0&0\\
\Phi_{1,1}&\Phi_{2,1}&\la_1\Phi_{1,1}&\la_1\Phi_{2,1}&\la_1^2\Phi_{2,1}\\
\Phi_{1,2}&\Phi_{2,2}&\la_2\Phi_{1,2}&\la_2\Phi_{2,2}&\la_2^2\Phi_{2,2}\\
\Phi_{1,3}&\Phi_{2,3}&\la_3\Phi_{1,3}&\la_3\Phi_{2,3}&\la_3^2\Phi_{2,3}\\
\Phi_{1,4}&\Phi_{2,4}&\la_4\Phi_{1,4}&\la_4\Phi_{2,4}&\la_4^2\Phi_{2,4}
\end{matrix}\right|,\ \ (\T_2)_{22}=\left|\begin{matrix}0&1&0&\la&\la^2\\
\Phi_{1,1}&\Phi_{2,1}&\la_1\Phi_{1,1}&\la_1\Phi_{2,1}&\la_1^2\Phi_{2,1}\\
\Phi_{1,2}&\Phi_{2,2}&\la_2\Phi_{1,2}&\la_2\Phi_{2,2}&\la_2^2\Phi_{2,2}\\
\Phi_{1,3}&\Phi_{2,3}&\la_3\Phi_{1,3}&\la_3\Phi_{2,3}&\la_3^2\Phi_{2,3}\\
\Phi_{1,4}&\Phi_{2,4}&\la_4\Phi_{1,4}&\la_4\Phi_{2,4}&\la_4^2\Phi_{2,4}
\end{matrix}\right|.\]

\end{theorem}
We can find
\[T_2(\la;\la_1,\la_2,\la_3,\la_4)|_{\la=\la_i}\left(\begin{matrix}\Phi_{1,i}\\
\Phi_{2,i}
\end{matrix}\right)=0\]
where $i=1,2,3,4.$
Similarly, for transformation $T_2$, following transformation formula holds
\[
T_{2t}+T_2U&=&U^{[2]}T_2,\\
T_{2z}+T_2V&=&V^{[2]}T_2,
\]
by which the relation between $E,p,\eta$ and $E^{[2]},p^{[2]},\eta^{[2]}$ will be got in the following  relation
\[\label{T2E2}U_{0}^{[2]}=U_0+[t_{1}^{[2]},\sigma_3],\]
\begin{eqnarray}\label{V[2]}
V^{[2]}_{-1}
&=&T_2|_{\la=-i\om b_2} V_{-1}T_2^{-1}|_{\la=-i\om b_2}.
\end{eqnarray}
This gives the relation between $E,p,\eta$ and $E^{[2]},p^{[2]},\eta^{[2]}$.
One can also get  following two-fold Darboux transformation in detail.

\[E^{[2]}&=&E-\frac{2}{b_1}(t_{1}^{[2]})_{12},\\
p^{[2]}&=&\frac{2\eta(T_2)_{11}(T_2)_{12}
-p^*(T_2)_{12}(T_2)_{12}+p(T_2)_{11}(T_2)_{11}}{(T_2)_{11}(T_2)_{22}-(T_2)_{12}(T_2)_{21}}|_{\la=-i\om b_2},\\
\eta^{[2]}&=&\frac{\eta((T_2)_{11}(T_2)_{22}+(T_2)_{12}(T_2)_{21})-p^*(T_2)_{12}(T_2)_{22}
+p(T_2)_{11}(T_2)_{21}}{(T_2)_{11}(T_2)_{22}-(T_2)_{12}(T_2)_{21}}|_{\la=-i\om b_2},
\]
where $(t_{1}^{[2]})_{12}$ is the element at the first row and second column in the matrix of  $t_{1}^{[2]}$.
This transformation will be used to generate two-soliton solutions of the IH-MB equation later.

As an application of these determinant representation of Darboux transformations  of IH-MB equations, soliton solutions and positon solutions will be constructed in the next section.

\sectionnew{Soliton solutions of the IH-MB equations}

In this section, first, we will consider the construction of one soliton solution of the IH-MB equations with suitable seed solutions.
Bring trivial seed solutions as $E=0, p=0, \eta=1$ into linear equations eqs.(\ref{linear}-\ref{linear2})., then the linear equations become

\[
\Phi_t&=&U\Phi,\\
\Phi_z&=&V\Phi,
\]
where
\[
\Phi&=&\left(\begin{matrix}\Phi_1\\ \Phi_2
\end{matrix}\right),\\
U&=&\left(\begin{matrix}\la& 0\\ 0& -\la
\end{matrix}\right),\\
V&=&\left(\begin{matrix}-4\lambda^3a_4-2i\lambda^2a_3-\lambda a_1-\frac{b_1a_7}{2\lambda+2i\omega b_2} & 0\\ 0& 4\lambda^3a_4+2i\lambda^2a_3+\lambda a_1+\frac{b_1a_7}{2\lambda+2i\omega b_2}
\end{matrix}\right).\]
Easy calculation can lead to following eigenfunctions
\[
\Phi_1&=&\exp(\la t+\int-4\lambda^3a_4-2i\lambda^2a_3-\lambda a_1-\frac{b_1a_7}{2\lambda+2i\omega b_2}dz+\frac{x_0+iy_0}2),\\
\Phi_2&=&\exp(-\la t+\int4\lambda^3a_4+2i\lambda^2a_3+\lambda a_1+\frac{b_1a_7}{2\lambda+2i\omega b_2}dz-\frac{x_0+iy_0}2+i\theta),\]
where $x_0$,$y_0$ and $\theta$ are all arbitrary fixed
real constants. Substituting these two eigenfunctions into the one-fold Darboux transformation eq.\eqref{1darbouxE}, eq.\eqref{1darbouxp} and eq.\eqref{1darbouxy} and  choosing $\la=\alpha_1+\beta_1 i$, $x_0=0,y_0=0, \theta=0$, then the following  solition solution are obtained:

\begin{eqnarray*}E_1&=&\frac{2\alpha_1}{b_1}  e^{- \frac{Az}{2(\alpha_1+\beta_1 i+\omega b_2 i)} +2 i t \beta_1+\frac{ -\bar Az}{2(-\alpha_1+\beta_1 i+\omega b_2 i) }} sech(\frac{Az}{2(\alpha_1+\beta_1 i+\omega b_2 i) }-2 t \alpha_1+\frac{ -\bar Az}{2(-\alpha_1+\beta_1 i+\omega b_2 i) })
\end{eqnarray*}

\begin{eqnarray*}p_1&=& -\frac{\alpha_1 e^{2 i t \beta_1} }{( \beta_1+ \omega b_2)^2+ \alpha_1^2}(\alpha_1 e^{2 t \alpha_1-\frac{Az}{\alpha_1+\beta_1 i+\omega b_2 i}}-\alpha_1 e^{-2 t \alpha_1+\frac{-\bar Az}{-\alpha_1+\beta_1 i+\omega b_2 i}}
+\beta_1 e^{-2 t \alpha_1+\frac{-\bar Az}{-\alpha_1+\beta_1 i+\omega b_2 i}} i\\
&&+\beta_1 e^{2 t \alpha_1-\frac{Az}{\alpha_1+\beta_1 i+\omega b_2 i}} i+\omega b_2 e^{2 t \alpha_1-\frac{Az}{\alpha_1+\beta_1 i+\omega b_2 i}} i+\omega b_2 e^{-2 t \alpha_1+\frac{-\bar Az}{-\alpha_1+\beta_1 i+\omega b_2 i}} i)\\
&&sech^2(\frac{Az}{2(\alpha_1+\beta_1 i+\omega b_2 i)}-2 t \alpha_1+\frac{-\bar Az}{2(-\alpha_1+\beta_1 i+\omega b_2 i)}) ),
\end{eqnarray*}

\begin{eqnarray*}\eta_1&=& (( cosh^2(-\frac{Az}{2(\alpha_1+\beta_1 i+\omega b_2 i)}+2 t \alpha_1-\frac{-\bar Az}{2(-\alpha_1+\beta_1 i+\omega b_2 i)})-2 \frac{\alpha_1^2}{(\beta_1 +\omega b_2)^2+\alpha_1^2)})\\
&&sech^2(\frac{Az}{2(\alpha_1+\beta_1 i+\omega b_2 i)}-2 t \alpha_1+\frac{-\bar Az}{2(-\alpha_1+\beta_1 i+\omega b_2 i)}),
\end{eqnarray*}

where
\begin{eqnarray*}A:&=&8 a_4 \alpha_1^4+2 i a_1 \alpha_1 \omega b_2-32 i a_4 \alpha_1 \beta_1^3-48 a_4 \alpha_1^2 \beta_1^2-24 a_4 \alpha_1^2 \beta_1 \omega b_2-12 i a_3 \alpha_1 \beta_1^2\\
&&+32 i a_4 \alpha_1^3 \beta_1+8 a_4 \beta_1^4+8 a_4 \beta_1^3 \omega b_2+4 i a_3 \alpha_1^3-12 a_3 \alpha_1^2 \beta_1-4 a_3 \alpha_1^2 \omega b_2-24 i a_4 \alpha_1 \beta_1^2 \omega b_2\\
&&+8 i a_4 \alpha_1^3 \omega b_2+4 a_3 \beta_1^3+4 a_3 \beta_1^2 \omega b_2+2 a_1 \alpha_1^2+4 i a_1 \alpha_1 \beta_1-8 i a_3 \alpha_1 \beta_1 \omega b_2\\
&&-2 a_1 \beta_1^2-2 a_1 \beta_1 \omega b_2+b_1 a_7).\end{eqnarray*}

\begin{figure}[h!]
\centering
\raisebox{0.85in}{($|E|^2$)}\includegraphics[scale=0.18]{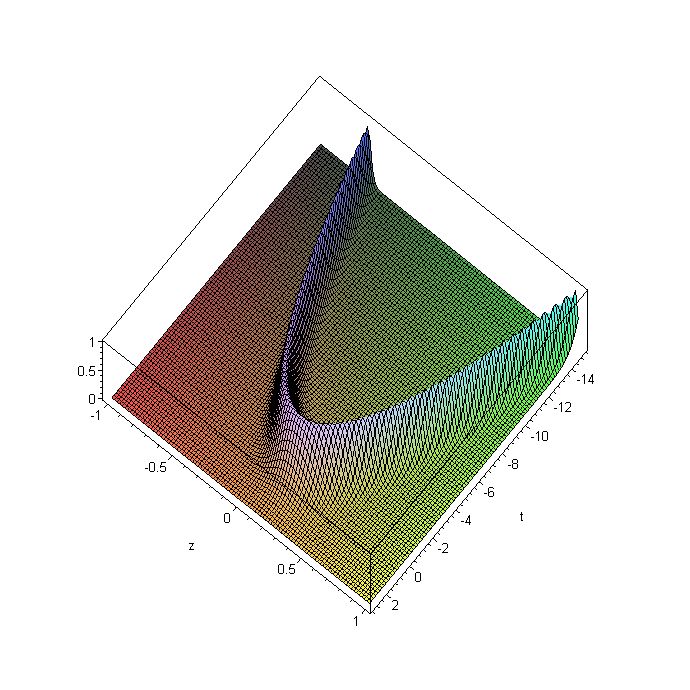}
\hskip 0.03cm
\raisebox{0.85in}{($|p|^2$)}\raisebox{-0.1cm}{\includegraphics[scale=0.18]{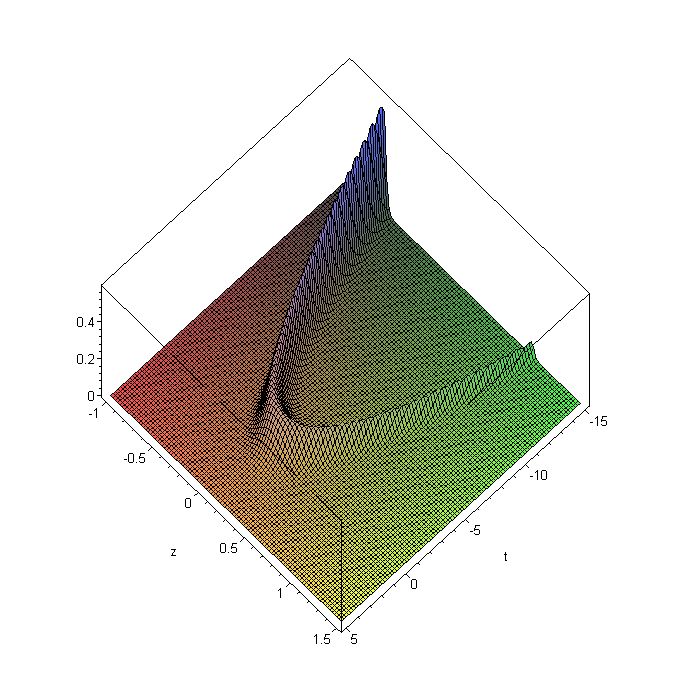}}
\hskip 0.03cm
\raisebox{0.85in}{($\eta$)}\raisebox{-0.1cm}{\includegraphics[scale=0.18]{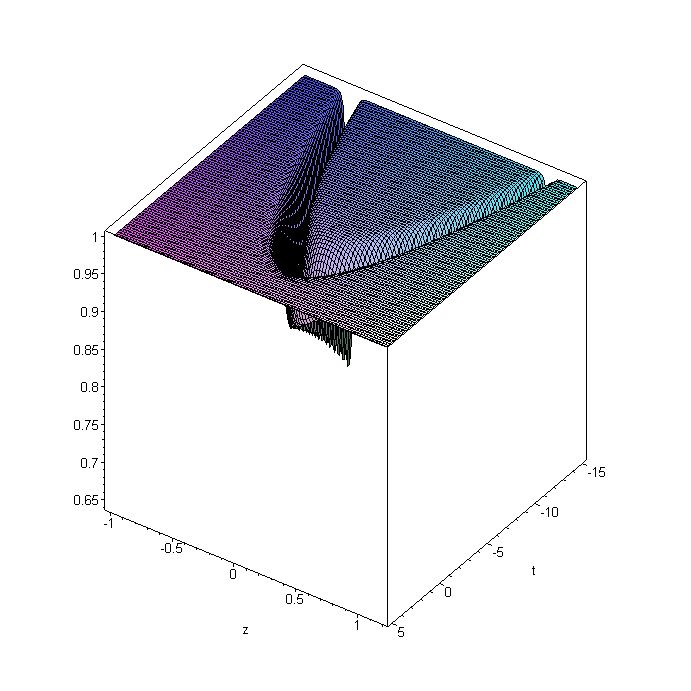}}
 \caption{One solition solution $(E,p,\eta)$  of the IH-MB equations when $a_1=z,a_3=-1,a_4=z,a_7=z,b_1=1,b_2=z,\omega=1.5,\alpha_1=0.5,\beta_1=2,$.}\label{1solitonH-MB}
\end{figure}

Similarly, substituting these two eigenfunctions into the one-fold Darboux transformation eq.\eqref{1darbouxE}, eq.\eqref{1darbouxp} and eq.\eqref{1darbouxy}, and  taking $a_1=z,a_3=-1,a_4=z,a_7=z,b_1=1,b_2=z,\omega=1.5,\alpha_1=0.5,\beta_1=2,$ then the one-solition solutions of the classical H-MB equations can be obtained whose evolution is given in Fig.\ref{1solitonH-MB}, which clearly indicates that $E$ and $p$ are bright solitons because their waves are under the flat non-vanishing plane whereas $\eta$ is a dark soliton.

Now let us discuss about the construction of the two-soliton solution of IH-MB system. For the purpose of construction of  two soliton solution, we need to use two spectral parameters  $\lambda_1=\alpha_1+\beta_1 i$
and  $\lambda_2=\alpha_2+\beta_2 i$. After the second Darboux transformation, we can construct the two solition solution.  As the general form of two soliton solution is tedious in nature, for simplicity, we will give only the two soliton solution of E  taking values as $a_1=z,a_3=-1,a_4=z,a_7=z,b_1=1,b_2=z,\omega=1.5,\alpha_1=0.5,\beta_1=2,\alpha_2=1,\beta_2=1.5$ in appendix.

\begin{figure}[h!]
\centering
\raisebox{0.85in}{($|E|^2$)}\includegraphics[scale=0.18]{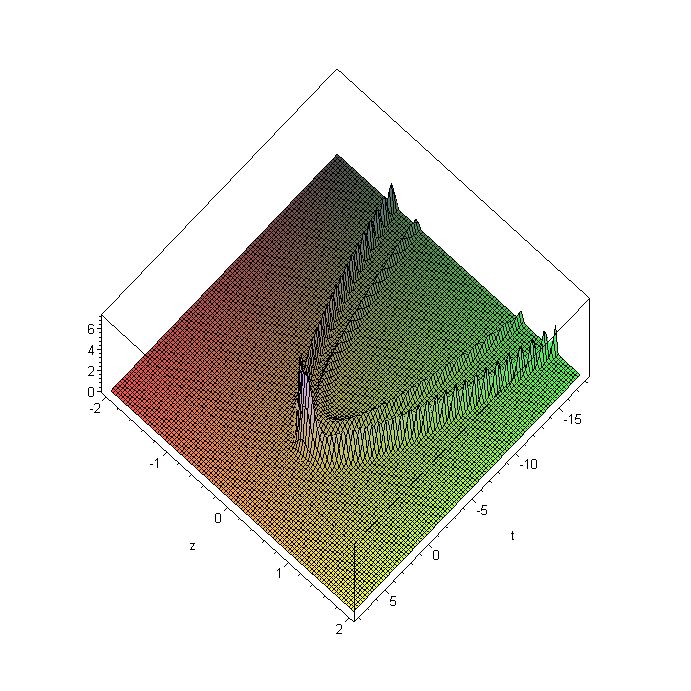}
\hskip 0.01cm
\raisebox{0.85in}{($|p|^2$)}\raisebox{-0.1cm}{\includegraphics[scale=0.23]{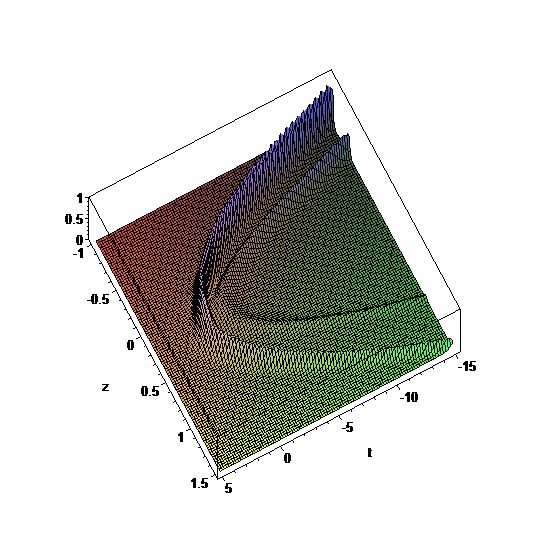}}
\hskip 0.01cm
\raisebox{0.85in}{($\eta$)}\raisebox{-0.1cm}{\includegraphics[scale=0.18]{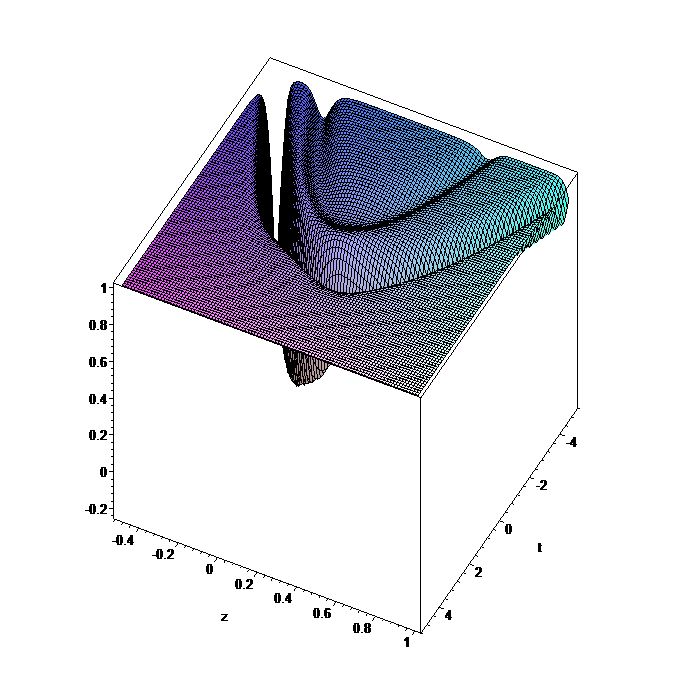}}
 \caption{Two solition solution $(E,p,\eta)$  of the IH-MB equations when $a_1=z,a_3=-1,a_4=z,a_7=z,b_1=1,b_2=z,\omega=1.5,\alpha_1=0.5,\beta_1=2,\alpha_2=1,\beta_2=1.5$.}\label{2solitonH-MB}
\end{figure}

We also construct  two soliton solutions for p and $\eta$ in a similar manner. For completeness, instead of giving complicated forms of p and $\eta$,  the graphical representation of them is shown in Fig.\ref{2solitonH-MB}.

If we suppose $a_1=0,a_2=0,a_3=-1,a_4=1,a_5=6,a_6=-2,a_7=-2,b_1=1,b_2=-1,$ this case will go to the classical H-MB equation\cite{rogueHMB} with constant coefficients.

\sectionnew{Bright and dark positon solutions of IH-MB system}

For the two soliton solution constructed in the last section, if the second spectral parameter $\la_2$ is close to the first spectral parameter $\la_1$, doing the Taylor expansion of wave function up to first order near $\la_1$ will lead to a new kind of solution. This is exactly the so-called  positon solutions. In this section,  the construction of positon solution of IH-MB equations will be given. Firstly, following four linear functions out of linear system will be used to construct the second Darboux transformation which further generate the positon solutions,

\[
\Phi_1&=&\exp(\la_1 t+\int-4\lambda_1^3a_4-2i\lambda_1^2a_3-\lambda_1 a_1-\frac{b_1a_7}{2\lambda_1+2i\omega b_2}dz+\frac{x_0+iy_0}2),\\
\Phi_2&=&\exp(-\la_1 t+\int4\lambda_1^3a_4+2i\lambda_1^2a_3+\lambda_1 a_1+\frac{b_1a_7}{2\lambda_1+2i\omega b_2}dz-\frac{x_0+iy_0}2+i\theta)\\
\Phi_3&=&\exp(\la_2 t+\int-4\lambda_2^3a_4-2i\lambda_2^2a_3-\lambda_2 a_1-\frac{b_1a_7}{2\lambda_2+2i\omega b_2}dz+\frac{x_0+iy_0}2),\\
\Phi_4&=&\exp(-\la_2 t+\int4\lambda_2^3a_4+2i\lambda_2^2a_3+\lambda_2 a_1+\frac{b_1a_7}{2\lambda_2+2i\omega b_2}dz-\frac{x_0+iy_0}2+i\theta).\]
Then we define the following functions as
\[
&&\Phi_{1,1}:=\phi_1;\ \Phi_{1,2}:=-\phi_2^*;\ \Phi_{2,1}:=\phi_2;\ \Phi_{2,2}:=\phi_1^*;\\
&&\Phi_{1,3}:=\phi_3;\ \Phi_{1,4}:=-\phi_4^*;\ \Phi_{2,3}:=\phi_4;\ \Phi_{2,4}:=\phi_3^*.
\]

Now we take $\la_2=\la_1+\ep(1+i)$ and using the  Taylor expansion of wave function $\phi_3$ and $\phi_4$ up to first order of $\ep$ in terms of $\la_1$.  Substitution of these manipulations into the second Darboux transformation discussed in the last section will lead to positon solutions. For example, after taking values as $a_1=0.5,a_3=z,a_4=z,a_7=1,b_1=1,b_2=1,\omega=1.5,\alpha_1=0.5,\beta_1=1,$ i.e. the case of classical H-MB equations\cite{rogueHMB}, the positon solutions $(E_{p},p_{p},\eta_{p})$  can be derived. Here for simplicity, we only give positon solution $E_{p}$ in following form

\begin{eqnarray*}E_{p}&=& -0.2e^{2 i t+2.5 i z^2-0.6153846154 i z} [(-130000000000 z^2+140000000000 i z^2+11420118340 z\\
&&+591715977 i z-20000000000 t-20000000000) e^{-t-7.500000000 z^2+0.5769230768 z}\\
&&+(591715977i z +140000000000 i z^2 +130000000000 z^2 -11420118340 z +20000000000 t\\
&&  -20000000000 ) e^{t+7.500000000 z^2-0.5769230768 z}]\\
&&/(2000000000 cosh(-15z^2+1.153846154 z-2t)+4000000000 t^2+52000000010 t z^2\\
&&+2000000000-4568047338 t z+365000000000 z^4\\
&&-28035502960 z^3+1307692307z^2).
\end{eqnarray*}

\begin{figure}[h!]
\centering
\raisebox{0.85in}{($|E|^2$)}\includegraphics[scale=0.19]{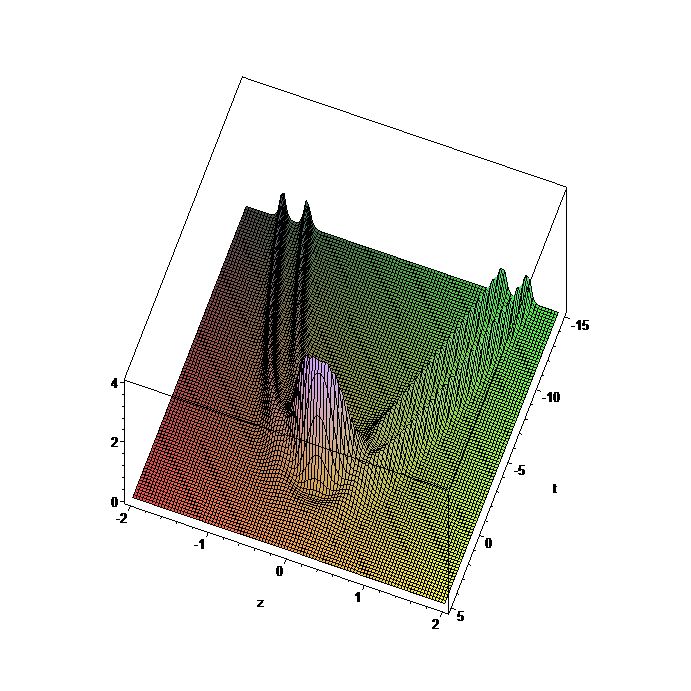}
\hskip 0.03cm
\raisebox{0.85in}{($|p|^2$)}\raisebox{-0.1cm}{\includegraphics[scale=0.19]{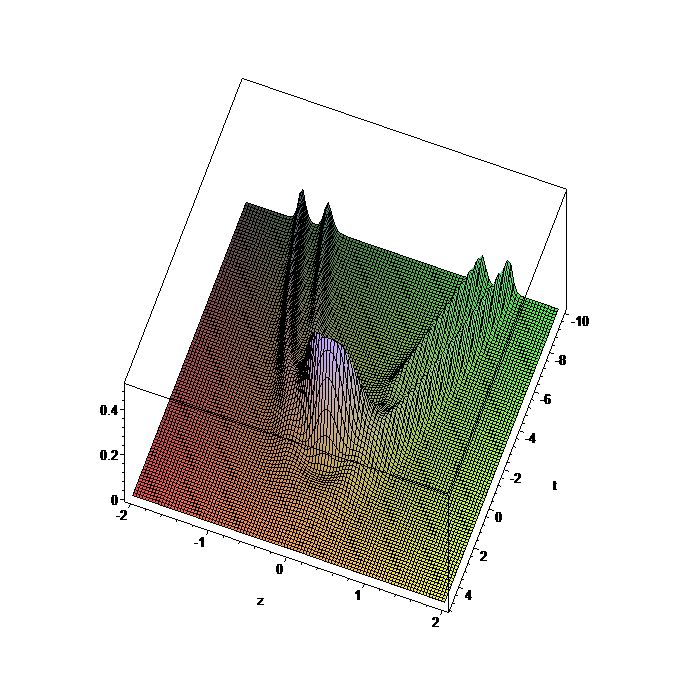}}
\hskip 0.03cm
\raisebox{0.85in}{($\eta$)}\raisebox{-0.1cm}{\includegraphics[scale=0.21]{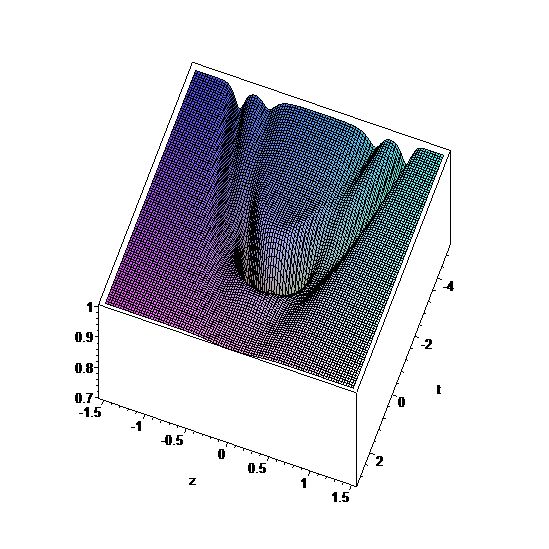}}
 \caption{One positon solution $(E,p,\eta)$  of the IH-MB equations when $a_1=0.5,a_3=z,a_4=z,a_7=1,b_1=1,b_2=1,\omega=1.5,\alpha_1=0.5,\beta_1=1.$}\label{Hpositon}
\end{figure}

\begin{figure}[h!]
\centering
\raisebox{0.85in}{($|E|^2$)}\includegraphics[scale=0.21]{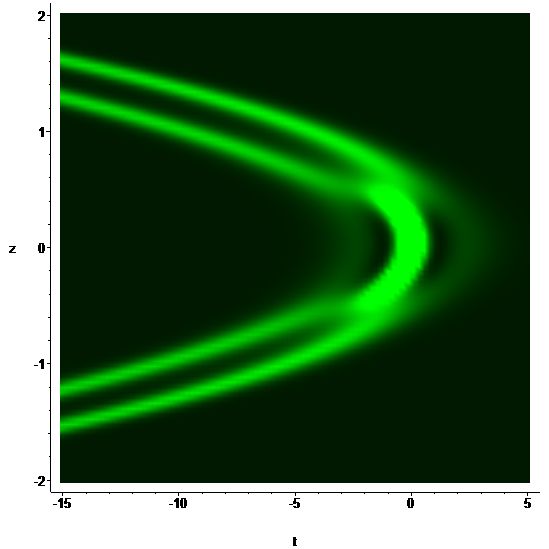}
\hskip 0.03cm
\raisebox{0.85in}{($|p|^2$)}\raisebox{-0.1cm}{\includegraphics[scale=0.17]{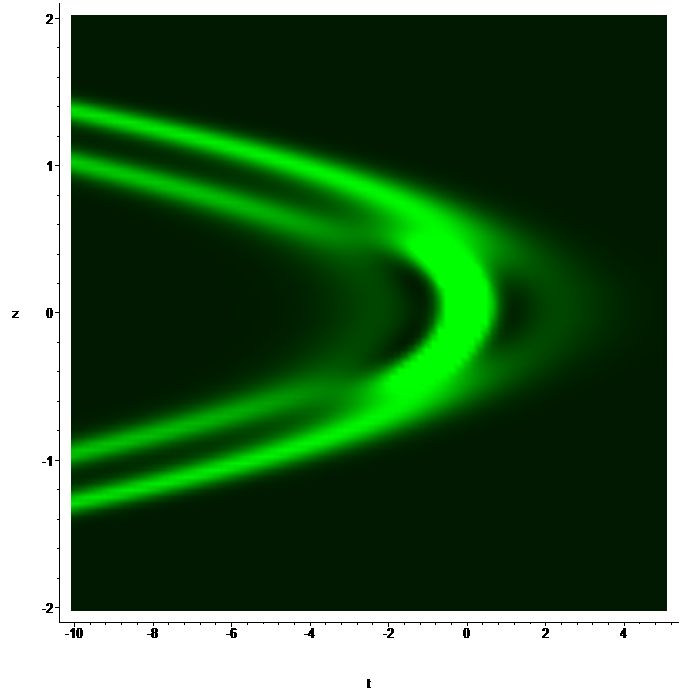}}
\hskip 0.03cm
\raisebox{0.85in}{($\eta$)}\raisebox{-0.1cm}{\includegraphics[scale=0.17]{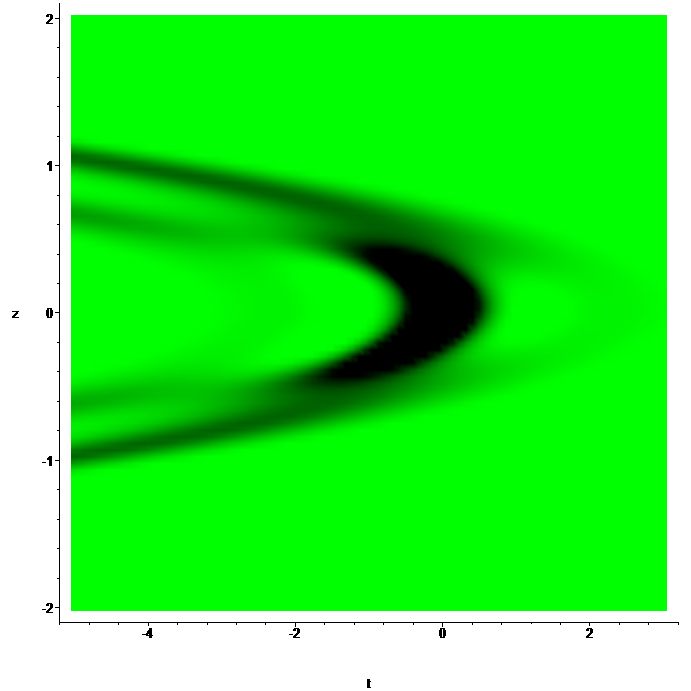}}
 \caption{One positon solution $(E,p,\eta)$  of the IH-MB equations when $a_1=0.5,a_3=z,a_4=z,a_7=1,b_1=1,b_2=1,\omega=1.5,\alpha_1=0.5,\beta_1=1.$}\label{Hpositonzdenstiy}
\end{figure}

\begin{figure}[h!]
\centering
\raisebox{0.85in}{($|E|^2$)}\includegraphics[scale=0.20]{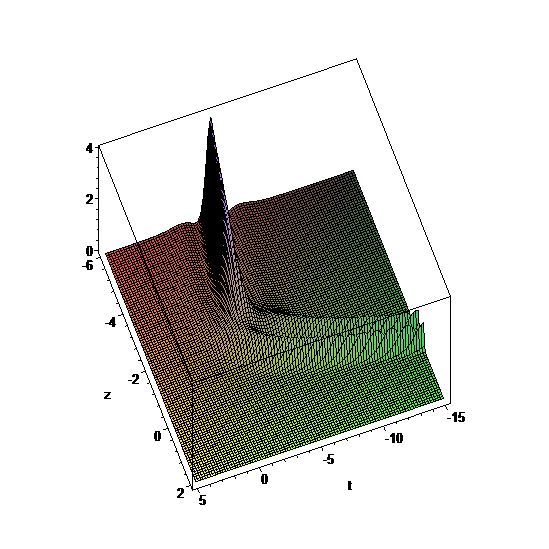}
\hskip 0.03cm
\raisebox{0.85in}{($|p|^2$)}\raisebox{-0.1cm}{\includegraphics[scale=0.18]{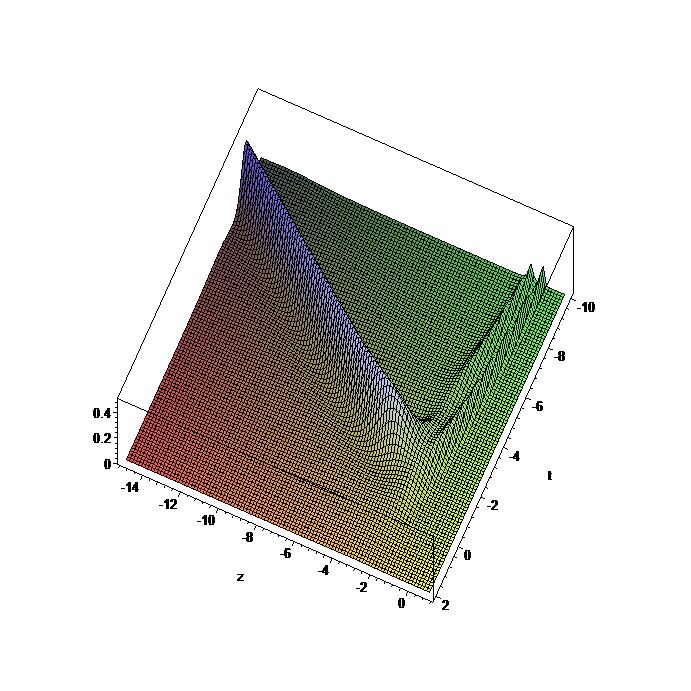}}
\hskip 0.03cm
\raisebox{0.85in}{($\eta$)}\raisebox{-0.1cm}{\includegraphics[scale=0.18]{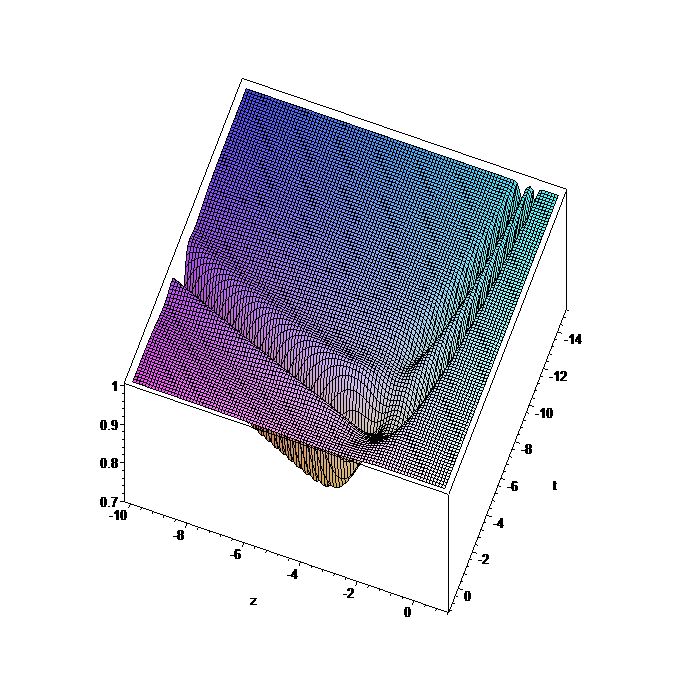}}
 \caption{One positon solution $(E,p,\eta)$  of the IH-MB equations when $a_1=0.5,a_3=e^z,a_4=e^z,a_7=1,b_1=1,b_2=1,\omega=1.5,\alpha_1=0.5,\beta_1=1$.}\label{positonexp}.
\end{figure}

\begin{figure}[h!]
\centering
\raisebox{0.85in}{($|E|^2$)}\includegraphics[scale=0.18]{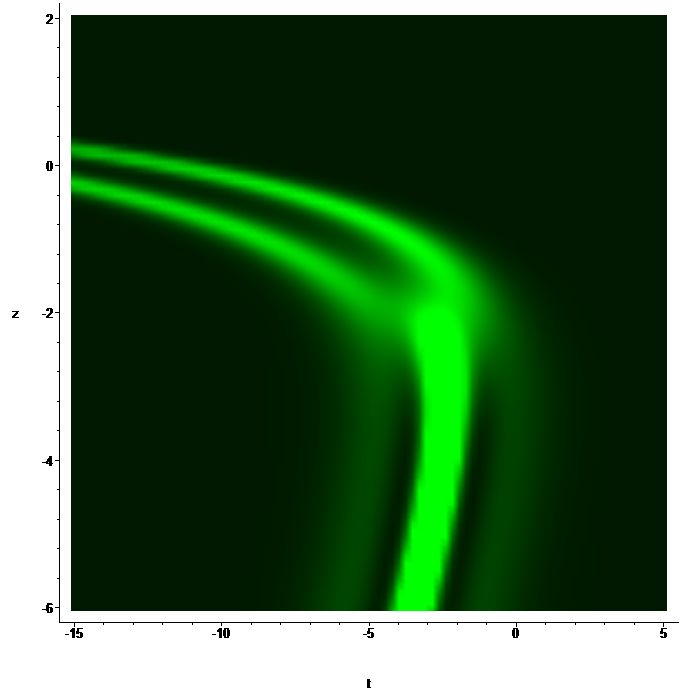}
\hskip 0.03cm
\raisebox{0.85in}{($|p|^2$)}\raisebox{-0.1cm}{\includegraphics[scale=0.18]{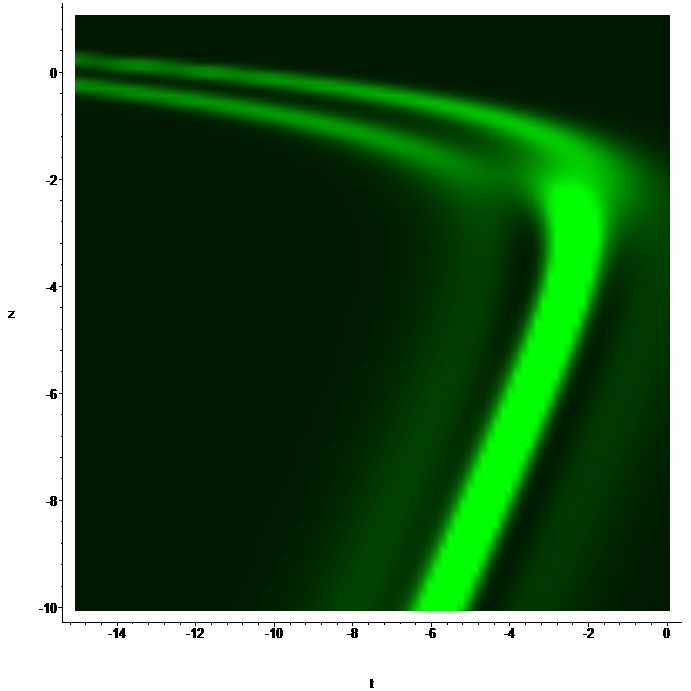}}
\hskip 0.03cm
\raisebox{0.85in}{($\eta$)}\raisebox{-0.1cm}{\includegraphics[scale=0.18]{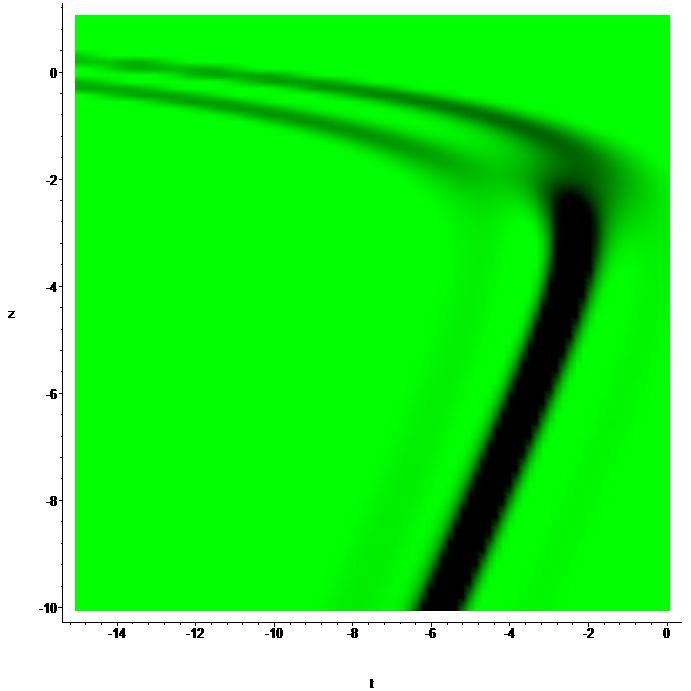}}
 \caption{One positon solution $(E,p,\eta)$  of the IH-MB equations when $a_1=0.5,a_3=e^z,a_4=e^z,a_7=1,b_1=1,b_2=1,\omega=1.5,\alpha_1=0.5,\beta_1=1$.}\label{positonexpdensity}.
\end{figure}

The pictorial representation of positon solutions of the IH-MB equations is shown in Fig.\ref{Hpositon},

whose density plot is as Fig.\ref{Hpositonzdenstiy}.
Next, after taking values as $a_1=0.5,a_3=e^z,a_4=e^z,a_7=1,b_1=1,b_2=1,\omega=1.5,\alpha_1=0.5,\beta_1=1$,the picture of positon solutions of the IH-MB equations is as Fig.\ref{positonexp},
whose density plot is as Fig.\ref{positonexpdensity}.

From above we find that $E$ and $p$ are bright positon solutions whereas $\eta$ is a dark positon. In a similar way, using the higher order Darboux transformation,
one can also generate higher-order bright and dark positon solutions which will be omitted here. These positons with variable coefficients are different from the
classical H-MB equations which can be seen from their graphs.

\newpage

\sectionnew{Conclusion and Discussions}

In this paper, we derived the Darboux transformation of the inhomogeneous Hirota and the Maxwell-Bloch(IH-MB) equations governed by ultra-short pulse propagation through erbium doped optical waveguide.  Further  matrix representation of Darboux transformation of this system is constructed.  As examples, soliton solutions, positon solutions of the IH-MB equations have been constructed explicitly by using  Darboux transformation from trivial solutions seed solutions.   There are a few unclear interesting questions such as  the physical interpretations and observation of higher-order positon solutions, rogue waves solutions and their applications in physics?

{\bf Acknowledgments} {\noindent \small  This work  is supported by
NSF of Zhejiang Province under Grant No. LY12A01007, the NSF of China under Grant  No.11201251, 10971109, K.C.Wong Magna Fund in
Ningbo University and Program for NCET
under Grant No.NCET-08-0515. }

\newpage{}

\sectionnew{Appendix}

\begin{eqnarray*}&&E^{[1]}_{2-sol}\\
&&=[8(17+24z+9z^2)^{\frac29}(13+18z+9z^2)^{\frac29-\frac13i}e^{8z-t-\frac89arctan(4+3z)-23z^2+i(3t-\frac{13}{3}z-6z^2)}
(\frac{1+4i+3iz}{\sqrt{17+24z+9z^2}})^
{\frac29}\\
&&+8(13+18z+9z^2)^{\frac29-\frac13i}e^{-8z+t+3it+\frac89arctan(4+3z)+23z^2-\frac{13}{3}iz-6iz^2}(\frac{1+4i+3iz}{\sqrt{17+24z+9z^2}})^
{\frac29}\\
&&-2(13+18z+9z^2)^{\frac49}(17+24z+9z^2)^{\frac19-\frac49i}e^{12z-2t+4it-\frac23arctan(\frac32+\frac32z)-22z^2-\frac{43}3iz+24iz^2}(\frac{2+3i+3iz}{\sqrt{13+18z+9z^2}})^{\frac49}\\
&&-2(17+24z+9z^2)^{\frac19-\frac49i}e^{-12z+2t+4it+\frac23arctan(\frac32+\frac32z)+22z^2-\frac{43}3iz+24iz^2}(\frac{2+3i+3iz}{\sqrt{13+18z+9z^2}})^{\frac49}\\
&&
+4i(17+24z+9z^2)^{\frac29}(13+18z+9z^2)^{\frac29-\frac13i}e^{8z-t+3it-\frac89arctan(4+3z)-23z^2-13/3iz-6iz^2}
(\frac{1+4i+3iz}{\sqrt{17+24z+9z^2}})^
{\frac29}\\
&&+4i(17+24z+9z^2)^(\frac19-\frac49i)e^{-12z+2t+4it+\frac23arctan(\frac32+\frac32z)+22z^2-\frac{43}3iz+24iz^2}(\frac{2+3i+3iz}{\sqrt{13+18z+9z^2}})^{\frac49}\\
&&-4i(13+18z+9z^2)^(\frac29-\frac13i)e^{-8z+t+3it+\frac89arctan(4+3z)+23z^2-13/3iz-6iz^2}
(\frac{1+4i+3iz}{\sqrt{17+24z+9z^2}})^
{\frac29}\\
&&-4i(13+18z+9z^2)^{\frac49}(17+24z+9z^2)^{\frac19-\frac49i}e^{12z-2t+4it-\frac23arctan(\frac32+\frac32z)-22z^2-\frac{43}3iz+24iz^2}
(\frac{2+3i+3iz}{\sqrt{13+18z+9z^2}})^{\frac49}]\\
&&/[(\frac{2+3i+3iz}{\sqrt{13+18z+9z^2}})^{\frac49}(\frac{1+4i+3iz}{\sqrt{17+24z+9z^2}})^
{\frac29}(-8(13+18z+9z^2)^{\frac29}(17+24z+9z^2)^{\frac19}cos\\
&&(t+\frac49arctan(\frac32+\frac32z)+\frac13ln(13+18z+9z^2)
-10z+30z^2-\frac29arctan(4+3z)-\frac49ln(17+24z+9z^2))\\
&&+5(17+24z+9z^2)^{\frac29}e^{-4z+t+\frac23arctan(\frac32+\frac32z)
-\frac89arctan(4+3z)-z^2}+e^{-20z+3t+\frac23arctan(\frac32+\frac32z)+\frac89arctan(4+3z)+45z^2}\\
&&+5(13+18z+9z^2)^{\frac49}
e^{4z-t-\frac23arctan(\frac32+\frac32z)+\frac89arctan(4+3z)+z^2}\\
&&+(13+18z+9z^2)^{\frac49}(17+24z+9z^2)^{\frac29}e^{20z-3t-\frac23arctan(\frac32+\frac32z)-\frac89arctan(4+3z)-45z^2})].
\end{eqnarray*}



\end{document}